\pdfoutput=1
\documentclass[aps,prd,amsmath,floats,floatfix,superscriptaddress,nofootinbib,showpacs]{revtex4}

\usepackage[T1]{fontenc}
\usepackage[utf8]{inputenc}
\usepackage{lmodern}
\usepackage{lipsum}

\usepackage[dvipsnames, usenames]{xcolor}
\definecolor{linkcolor}{rgb}{0.0,0.3,0.5}
\usepackage[hypertexnames=false, unicode, colorlinks=true, linkcolor=linkcolor,
citecolor=linkcolor, filecolor=linkcolor,urlcolor=linkcolor,
pdfusetitle]{hyperref}

\usepackage[all]{hypcap}
\usepackage{graphicx}
\usepackage{xspace}
\usepackage{amssymb}
\usepackage[normalem]{ulem} 
\usepackage{bm} 
\usepackage{appendix}

\usepackage{microtype}

\usepackage[english]{babel}
\usepackage{blindtext}

\graphicspath{
  {figs/}
}

\usepackage{tikz}
\usetikzlibrary{shapes,snakes}
\usetikzlibrary{arrows,intersections,patterns,decorations.markings,shapes.geometric}
\usetikzlibrary{calc,fadings,decorations.pathreplacing,positioning}
\usetikzlibrary{decorations.pathmorphing} 
\tikzset{snake it/.style={decorate, decoration=snake}}

\usepackage{pgfplots}
\usepgfplotslibrary{colormaps}
\usetikzlibrary{intersections,patterns,pgfplots.fillbetween}

\tikzset{->-/.style={decoration={
  markings,
  mark=at position .5 with {\arrow{>}}},postaction={decorate}}
}
\tikzset{-<-/.style={decoration={
  markings,
  mark=at position .5 with {\arrow{<}}},postaction={decorate}}
}

\tikzset{
  >=latex, 
  inner sep=0pt,
  outer sep=2pt,
  mark coordinate/.style={inner sep=0pt,outer sep=0pt,minimum size=3pt,
    fill=black,circle}
}

\DeclareMathAlphabet{\mathpzc}{OT1}{pzc}{m}{it}

\renewcommand{\vec}[1] {\bm{#1}}

\newcommand*{\df}  {\delta}

\newcommand*{\non} {\nonumber}
\newcommand*{\lb} {\left(}
\newcommand*{\rb} {\right)}

\newcommand*{\la} {\left\langle}
\newcommand*{\ra} {\right\rangle}

\newcommand{\eq}[1]{\begin{align}#1\end{align}}
\newcommand{\eeq}[1]{\begin{equation}#1\end{equation}}

\begin{document}

\rightline{\scriptsize RBI-ThPhys-2023-15}

\title{On the asymptotic connection between full- and flat-sky angular correlators}

\newcommand\alvisehome{
\affiliation{Dipartimento di Fisica Galileo Galilei, Universit\` a di Padova, I-35131 Padova, Italy,}
\affiliation{INFN Sezione di Padova, I-35131 Padova, Italy,}
\affiliation{INAF-Osservatorio Astronomico di Padova, Italy,}
\affiliation{Theoretical Physics Department, CERN, 1 Esplanade des Particules, 1211 Geneva 23, Switzerland,}
}
\newcommand\zvonehome{
\affiliation{Ru\dj er Bo\v{s}kovi\'c Institute, Bijeni\v{c}ka cesta 54, 10000 Zagreb, Croatia,}
\affiliation{Kavli Institute for Cosmology, University of Cambridge, Cambridge CB3 0HA, UK,}
\affiliation{Department of Applied Mathematics and Theoretical Physics, University of Cambridge, Cambridge CB3 0WA, UK.}
}
\newcommand\zuchenghome{
\affiliation{Kavli Institute for Cosmology, University of Cambridge, Cambridge CB3 0HA, UK,}
\affiliation{Institute of Astronomy, University of Cambridge, Cambridge CB3 0HA, UK,}
}

\author{Zucheng Gao}
\email{zg285@cam.ac.uk}
\zuchenghome

\author{Alvise Raccanelli}
\email{alvise.raccanelli.1@unipd.it}
\alvisehome

\author{Zvonimir Vlah}
\email{zvlah@irb.hr}
\zvonehome

\begin{abstract}
We investigate the connection between the full- and flat-sky angular power spectra. First, we revisit this connection established on the geometric and physical grounds, namely that the angular correlations on the sphere and in the plane (flat-sky approximation) correspond to each other in the limiting case of small angles and a distant observer. To establish the formal conditions for this limit, we first resort to a simplified shape of the 3D power spectrum, which allows us to obtain analytic results for both the full- and flat-sky angular power spectra. Using a saddle-point approximation, we find that the flat-sky results are obtained in the limit when the comoving distance and wave modes $\ell$ approach infinity at the same rate. This allows us to obtain an analogous asymptotic expansion of the full-sky angular power spectrum for general 3D power spectrum shapes, including the $\Lambda$CDM Universe. In this way, we find a robust limit of correspondence between the full- and flat-sky results. These results also establish a mathematical relation, i.e.,~an asymptotic expansion of the ordinary hypergeometric function of a particular choice of arguments that physically corresponds to the flat-sky approximation of a distant observer. This asymptotic form of the ordinary hypergeometric function is obtained in two ways: relying on our saddle-point approximation and using some of the known properties of the hypergeometric function.
\end{abstract}

\maketitle

\section{Introduction}
\label{sec:intro}

The angular power spectrum is a powerful tool for analysing data from cosmological surveys. It is the canonical observable in the study of the distribution of temperature and polarisation anisotropies in the cosmic microwave background (CMB), as well as one of the possibilities when analyzing the distribution of matter (and its tracers) in the study of the large-scale structure (LSS) of the universe. The angular power spectrum is calculated by decomposing the CMB or matter tracer observables into a series of eigenfunctions that describe how the temperature or density of the universe varies with the direction on the sky. The angular power spectrum measures how much power is present in each spherical harmonic and how that power is distributed over different angular scales.

Two typical methods for calculating the angular power spectrum are the full-sky approach and the flat-sky approximation. In the full-sky approach, the entire spherical geometry of the sky is considered, with the observer located at the centre of the sphere (neglecting space-time curvature). The eigenfunctions on a sphere are simple spherical harmonics, so the angular power spectrum is thus a measure of the power in each of these harmonics. While the full-sky approach is well suited for analysing data from experiments that observe the entire sky (e.g.,~Planck~\cite{Planck++:2018}), the flat-sky approximation is often used for ground-based experiments that observe a smaller region of the sky (until recently, this has been a typical setup for galaxy surveys). Usually it is also assumed that the observations lie on a single plane in the sky, neglecting correlations along the line of sight. The latter, paired with the flat-sky geometry, forms the so-called Limber approximation~\cite{Limber:54, Kaiser:1984}; a practical `go-to' implementation of the angular power spectrum for LSS data analyses. However, some of the upcoming and planned cosmological LSS surveys, such as e.g.,~Euclid~\cite{Euclid:2011}, DESI~\cite{DESI:2016}, SPHEREx~\cite{Dore:2014}, SKAO~\cite{Abdalla:2015}, the Vera Rubin Observatory~\cite{Abell:2009}, will observe large portions of the sky. This means that the approximations currently in use can no longer provide sufficient accuracy and that upcoming data analyses will need to go beyond existing practices (see, e.g.,~\cite{Leonard:2022} for a recent comparison of various methods).

On the other hand, solely relying on the full-sky formalism would make the analysis pipeline  cumbersome and even prevent efficient information extraction and usage of entire data sets, as we would be forced to perform various suboptimal data compressions and binning. Constrained by these two considerations, the optimal strategy is thus to find a middle path by removing some limitations of the currently implemented approximations while retaining most of the computational simplicity and efficiency of such approximations. In this paper, we lay out the map of this intermediate path by providing a consistent derivation of a new flat-sky result as an asymptotic approximation of the full-sky formalism.

This paper is organised as follows:
\begin{itemize}
\item Sec.~\ref{sec:3D_PS} provides a preamble to the discussion of the cosmological correlators and their projections on the sky. It introduces the theoretical unequal-time 3D power spectrum as a two-point correlation function in Fourier space over the statistical ensemble and, thus, by construction, a non-observable quantity (see~\cite{Raccanelli+:2023I} for a discussion).

\item Sec.~\ref{sec:unequal_time} introduces the full- and flat-sky two-point angular power spectrum. We discuss the relations between these spectra and establish their correspondence in the flat-sky limit. We discuss the emergence and consequences of unequal-time effects in the flat-sky two-point angular power spectrum. We show how these effects lead to the breaking of translational invariance in the 2D plane and, consequently, to the breakdown of the isotropy manifested in the full-sky angular power spectrum (see also~\cite{Raccanelli+:2023II}). 

\item In Sec.~\ref{sec:flat_sky_toy}, we continue our study of the angular power spectrum of the full- and flat-sky, starting from a simple analytic form of the theoretical 3D power spectrum. In this way, we obtain analytic expressions for both the full and flat-sky angular power spectra, allowing us to determine the precise asymptotic limit under which flat-sky results are obtained. Although these conditions were obtained in this simplified scenario, they can be generalized for arbitrary cosmologies and power spectra, including the  $\Lambda$CDM case.

\item Sec.~\ref{sec:flat_sky_limit_general} utilises the precise asymptotic limit conditions obtained in the previous section to derive the expression for the unequal-time angular power spectrum in flat-sky approximation for a general theoretical 3D power spectrum. For this purpose, we use the Mellin integral transform of the 3D power spectrum.

\item In Sec.~\ref{sec:math_result} we determine a limit of the ordinary hypergeometric function ${}_2F_1(a,b;c;z)$ in which it corresponds to the modified Bessel function of the second kind $K_\nu(z)$.  We achieve this by combining the known analytic solutions for the integral containing the power law and two spherical Bessel functions with our results from the previous sections. We also show an alternative derivation that follows from some of the known properties of the  ordinary hypergeometric function.

\item We end by summarising our results and providing some concluding remarks in Sec.~\ref{sec:conclusion}.
\end{itemize}

\begin{table*}
\centering
\begin{tabular}{l l}
\hline
\hline
$\df^{\rm K}_{ij}$ &~ Kronecker symbol \\ [1pt]
$\df^{\rm D} (\vec x)$ &~ Dirac delta function \\ [1pt]
$ W(\chi) $ &~ Window function; related to the specific observable and survey\\ [1pt]
\hline
$\df(\vec x)$ &~ 3D density field of matter or biased tracer \\ [1pt]
$\hat \df(\vec \theta)$ &~ 2D projected filed in the real space coordinates on the sky  \\ [1pt]
\hline
$\mathcal P(\vec k;\, z, z')$ &~ Unequal-time theoretical power spectrum of the 3D density field (unobservable) \\ [1pt]
$C_{\ell}$ &~ Projected angular power spectrum (with finite size window functions)  \\ [1pt]
$\mathbb{C}_{\ell}  (z,z')$ &~ Unequal-time angular power spectrum (in the narrow window function limit) \\ [1pt]
\hline
$J_\lambda (z)$ &~ ordinary Bessel function \\[1pt]
$j_\lambda (z)$ &~ spherical Bessel function \\[1pt]
$I_\lambda (z)$ &~ modified Bessel function of the first kind \\[1pt]
$K_\lambda (z)$ &~ modified Bessel function of the second kind \\[1pt]
${}_2F_1(a,b; c; z)$ &~ ordinary (Gaussian) hypergeometric function \\[1pt]
\hline
\hline
\end{tabular}
\caption{Notation used for the most important quantities in this paper.}
\label{tab:notation}
\end{table*}

Table~\ref{tab:notation} summarises the notation used throughout the paper for the most important physical and mathematical quantities.

\section{Theoretical 3D power spectrum}
\label{sec:3D_PS}

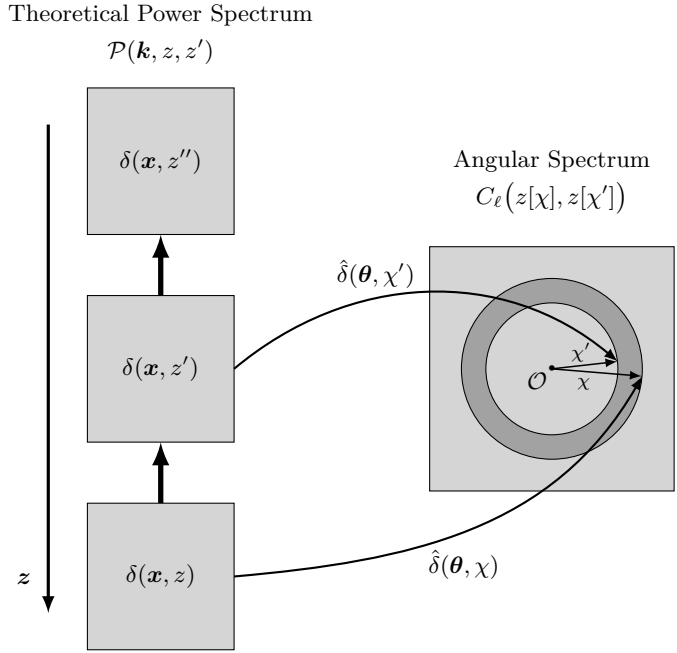
\begin{figure*}[t!]
\centering
\begin{tikzpicture}[scale=.65,every node/.style={minimum size=1cm},on grid]
                  
  \draw[<-, very thick] (-2.3,-5.0) -- (-2.3,5.0);
  \node[black, opacity = 0] at (3,-6) {\textbullet}; 
  
  \draw[->, line width=0.7mm] (0,-2.75) -- (0,-1.5);
  \draw[->, line width=0.7mm] (0,1.5) -- (0,2.75);
  
  \draw[fill=gray!50!white, draw=black, fill opacity = 0.65] (-1.5,2.75) rectangle (1.5, 5.75);
  \draw[fill=gray!50!white, draw=black, fill opacity = 0.65] (-1.5,-1.5) rectangle (1.5,1.5);
  \draw[fill=gray!50!white, draw=black, fill opacity = 0.65] (-1.5,-2.75) rectangle (1.5,-5.75);  

  \draw[fill=gray!50!white, draw=black, fill opacity = 0.65] (5.5,-2.5) rectangle (10.5,2.5);
  \draw[fill=gray!80!white, draw=black, fill opacity = 0.85, even odd rule] (8,0) circle (1.35) (8,0) circle (1.85);
  \node[black] at (8.0,0.0) {\scriptsize \textbullet}; 

  \draw [->,thick] (1.5,0.0) to [out=40,in=140] (9.35,0.15);
  \draw [->,thick] (1.5,-4.25) to [out=5,in=240] (9.85,-0.15);

  \draw[->, line width=0.2mm] (8.0, 0.0) -- (9.34,0.15);
  \draw[->, line width=0.2mm] (8.0, 0.0) -- (9.84,-0.15);  

  \node[black] at (0,4.25) {\small $\delta(\vec x, z'')$};    
  \node[black] at (0,0) {\small $\delta(\vec x, z')$};
  \node[black] at (0,-4.25) {\small $\delta(\vec x, z)$};
  \node[black] at (6.2,-4.0) {\small $\hat \delta(\bm \theta, \chi)$};  
  \node[black] at (4.4,1.95) {\small $\hat \delta(\bm \theta, \chi')$};  

  \node[black] at (7.7,-0.25) {\small $\mathcal O$};
  \node[black] at (8.65,-0.3) {\scriptsize $\chi$};
  \node[black] at (8.6,0.35) {\scriptsize $\chi'$};  

  \node[black] at (-2.8,-4.3) {${\bm z}$};

  \node[black] at (0.0,7.25) {Theoretical Power Spectrum};  
  \node[black] at (0.0,6.5) {\small $\mathcal P (\vec k, z, z')$};
  
  \node[black] at (8.0,4.25) {Angular Spectrum};  
  \node[black] at (8.0,3.5) {\small $  C_\ell \big(z[\chi], z[\chi'] \big)$};
\end{tikzpicture}
\caption{Scheme representing the construction of the observable angular power spectrum. We start by correlating the 3D density field $\delta(\vec x, z)$, which provides us with the theoretical, unobservable unequal-time 3D power spectrum $\mathcal P(\vec k; z,z')$. The simplest two-point observable accessible to an observer at position $\mathcal O$ is the unequal-time projected angular power spectrum $C_\ell(\chi,\chi')$.}
\label{fig:spectra}
\end{figure*}

The usual method for studying the dynamics of gravitational galaxy clustering uses 3D Fourier space correlators. The reason why Fourier space is an appropriate choice lies in the fact that the properties of the system, such as statistical isotropy and homogeneity, manifest themselves in the direct simplification of the functional form of the $n$-point correlation functions. The homogeneity manifests itself in the translational invariance of these correlators, while the isotropy corresponds to the rotational invariance. The realization of these properties in the correlation functions can be easily observed, e.g.,~in N-body simulations, where one can perform ensemble averaging over different realisations determined by different initial conditions (see~Fig.~\ref{fig:spectra}). The  \textit{theoretical 3D power spectrum} $\mathcal P(k)$ can then be defined as the two-point correlation function of the overdensity field of a tracer
\eeq{
\la \df(\vec k, z) \df(\vec k', z') \ra = (2\pi)^3 \delta^{\rm D}(\vec k + \vec k') \mathcal P(k;\, z, z') \, . 
}
Here we have explicitly pointed out that two overdensity fields do not have to be correlated at the same time, so we can study the unequal-time power spectrum. The left part of Fig.~\ref{fig:spectra} schematically shows a realisation of the time evolution of the overdensity. The right panel shows how these different times are projected onto a single observable redshift slice. Thus, the full unequal-time 3D power spectrum is not an observable quantity since it is never accessible from survey data to an observer located at a single position in the Universe (for more details, see~\cite{Raccanelli+:2023II}). On the other hand, the 2D angular power spectrum $C_\ell$ correlates the projected overdensity in two different redshift regions is the most easily observable quantity accessible to such an observer (see the sketch in the right part of Fig.~\ref{fig:spectra}).

The 3D power spectrum is characterised by its shape dependence (in wave modes $k$) as well as by the time dependence, which in turn are determined by the physical model and cosmological parameters of our universe. Thus, determining its shape and time dependence allows us to measure and constrain the fundamental parameters of our Universe. In this work, however, our goal is not to determine any of these parameters or to study their sensitivity in detail; more on that can be found in~\cite{Raccanelli+:2023I, Raccanelli+:2023II}. Rather, we explore the connections between the observable angular power spectrum and the unobservable 3D power spectrum in general terms, focusing on the broad properties of these relationships.  Although we present our final results in a general form so that they are also valid for the $\Lambda$CDM universe, in certain cases we will find it helpful to use a simple functional form that captures some general properties similar to those of the real universe. Thus, in Sec.~\ref{sec:flat_sky_toy}, we use $\mathcal P(k; z, z') = A D(z)D(z') k^2 \exp\lb- \alpha^2 k^2\rb$ to establish the asymptotic relation between the full- and flat-sky angular power spectra and to investigate the anti-correlations that appear in the unequal-time angular power spectrum.

\section{Unequal-time angular power spectrum}
\label{sec:unequal_time}

The relationships between the full-sky and the flat-sky angular power spectrum formalisms have been studied in great detail in the context of temperature fluctuations and polarisation~\cite{Seljak:1996, Zaldarriaga+:1996, Kamionkowski++:1996, Hu+:1997, Hu:2000, Lewis+:2006}. In this section, we revisit and review these results for scalar fields, emphasising the geometric aspects of the connection between the full-sky and flat-sky tracer number density. First, we verify that the formally introduced projected overdensity fields in the corresponding scales lead to the equivalent observable (a similar approach was taken in~\cite{Hu:2000}, which also motivated much of the discussion presented in this section). Namely, we focus on the angular power spectrum. We show that the derived observables in both cases match in an asymptotic sense, i.e.,~the flat-sky observable recovers some of the properties, such as statistical isotropy, in an approximate form. We show how the Limber approximation can be obtained from the flat-sky approximation, assuming wide window functions.

\begin{figure*}[t!]
\centering
\begin{tikzpicture}[scale=.8,every node/.style={minimum size=1cm},on grid]

  \draw[fill=gray!80!black, draw=black, fill opacity = 0.45] (6.9,-3.15) rectangle (7,3.15);  

  \coordinate (c1) at (-8,0);
  \draw[fill=gray!40!white, draw=black, fill opacity = 0.50]
  ($(c1) + (-12:14.9)$) arc (-12:12:14.9)
  --
  ($(c1) + (12:15.0)$) arc (12:-12:15.0)
  -- cycle;

  \node[black] at (3.35,3.05) {\small Full-sky surface};     
  \node[black] at (3.35,2.55) {\small Flat-sky surface};     
  \draw[<-, dashed, line width=0.1mm] (4.85,3.05) -- (6.61,3.05);
  \draw[<-, dashed, line width=0.1mm] (4.85,2.55) -- (6.975,2.55);

  \draw[->, line width=0.2mm] (-4.5,0) -- (-3,0);
  \draw[dotted, line width=0.3mm] (-2.95,0) -- (0.5,0);
  \draw[->, line width=0.2mm] (0.5,0) -- (6.9,0);

  \draw[->, line width=0.2mm] (-4.5,0) -- (-3.1,-0.375);
  \draw[dotted, line width=0.3mm] (-3.1,-0.375) -- (0.5,-1.325);
  \draw[->, line width=0.15mm] (0.5,-1.325) -- (6.6,-2.95);
  
  \draw[->, line width=0.15mm] (0.5,1.3) -- (6.9,1.3);
  \draw [dotted, line width=0.3mm] (-2.95,0) to [out=5,in=180] (0.5,1.3);

  \node[black] at (-4.5,0.0) {\small \textbullet}; 
  \node[black] at (-4.5,-0.3) {\small $\mathcal O$};  
  \node[black] at (7.25,2.85) {\small $\mathcal S$};

  \node[black] at (0.87,-1.10) {\small $ \chi'$};    
  \node[black] at (0.85,0.25) {\small $ \chi$};    
  \node[black] at (1.43,1.6) {\small $ \chi ( \hat n + \vec \theta )$};    

  \node[black] at (-3.75,0.35) {\small $\hat n$};

   \draw[<->, dashed, line width=0.15mm]  (7.35,-2.3) -- (7.35,2.3);   
  \node[black] at (7.6,0.0) {\small $\vec \theta$};    

\end{tikzpicture}
\caption{The full-sky and flat-sky geometrical setup. Assuming a limited survey volume and distant observer approximation, we expect the flat sky approximation to be a suitable representation of the full-sky results.}
\label{fig:full+flat_sky}
\end{figure*}
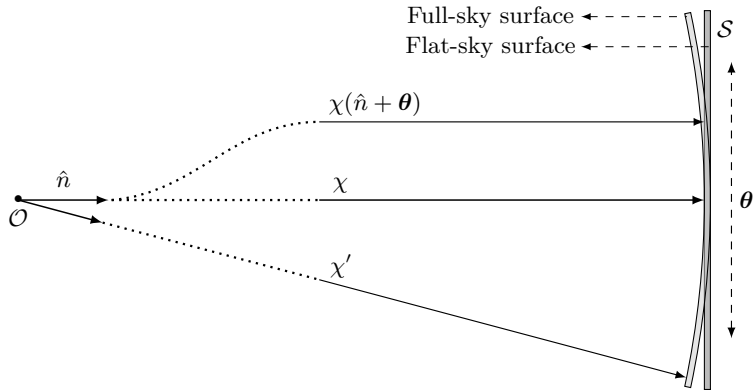

In Fig.~\ref{fig:full+flat_sky}, we schematically present our geometrical setup comparing the full-sky geometry to the flat-sky approximation. We can imagine a construction of the observable by collecting all the tracers in a certain redshift bin characterised by a isotropic window function $W(\chi)$ and a direction on the sky $\hat n$. We thus obtain a projected density field
\eeq{
 \hat \df (\hat n) = \int d\chi \;  W(\chi)  \df ( \chi \hat n, z[\chi]) 
 = \int d\chi \;  W(\chi)  \int \frac{d^3k}{(2\pi)^3} e^{- i \chi \hat n \cdot \vec k} \df ( \vec k, z[\chi]) \, ,
}
where the observable is obtained by projecting/integrating over the comoving distance $\chi$ weighted by the window function $W(\chi)$. On a full-sky, described by a spherical shell, it is convenient to represent the overdensity field in terms of the spherical harmonics expansion
\eeq{
 \hat \df ( \chi \hat n) = \sum_{\ell = 0}^\infty \sum_{m=-\ell}^\ell \hat\delta_{\ell,m}(\chi) Y_{\ell}^m(\hat n) \, .
}
This decomposition is useful as it allows us to utilise the statistical homogeneity and isotropy assumptions. The simplification is manifest once we look at the two-point statistics
\eeq{
\la \hat\delta_{\ell,m} \hat\delta_{\ell,m}^* \ra = \df^{\rm K}_{mm'} \df^{\rm K}_{\ell \ell'} C_{\ell} \, ,
}
where the two Kronecker delta functions $\df^{\rm K}_{mm'}$ and $\df^{\rm K}_{\ell \ell'}$ arise as consequences of the translational invariance and isotropy of the 3D power spectrum $\mathcal P(k)$. This allows us to introduce the \textit{projected angular power spectrum}, dependent on a single mode $\ell$, and is related to the 3D power spectrum $\mathcal P(k)$ via the well known relation 
\eeq{
C_{\ell} = 4\pi \int d \chi_1 d \chi_2\; W(\chi_1) W'(\chi_2) \int_0^\infty \frac{k^2 dk}{2\pi^2}\; 
\mathcal P(k; \chi_1,\chi_2) j_\ell(k \chi_1) j_\ell(k \chi_2)\, .
}
In recent years, there was a revival of efforts for efficient evaluations of this expression~\cite{Assassi++:2017, Schoneberg++:2018, Grasshorn++:2017, Fang++:2019, Chen++:2021, Feldbrugge:2023}.
If we are interested in thin redshift slices that characterise spectroscopic surveys, we have $W(\chi) = \df^{\rm D}\lb \chi - \chi^* \rb$, which simply gives us 
\eeq{
\label{eq:Cell_full}
\mathbb{C}^{\rm full}_{\ell}  (\chi,\chi')
= 4\pi \int_0^\infty \frac{k^2 dk}{2\pi^2}\; \mathcal P(k;\, \chi,\chi') \, j_\ell(k \chi) j_\ell(k \chi')\, ,
}
that we label the \textit{unequal-time angular power spectrum}. We use the explicit label `full' to distinguish the full-sky from the flat-sky version of the unequal-time angular power spectrum, which we investigate next. 

As indicated in Fig.~\ref{fig:full+flat_sky}, we can approximate the observable field near a given direction $\hat n$ and a given comoving distance $\chi$ by the flat-sky approximation rather than defining it on the spherical shell with the comoving distance $\chi$. This approximation assumes that all tracers lie in the same plane orthogonal to $\hat n$. Since the observer has a fixed location $\mathcal O$, this means that the statistical observable defined on such a plane is not guaranteed to inherit symmetries such as translational invariance in the plane. This can be recovered by explicitly assuming that a distant observer also implies a so-called \textit{plane-parallel approximation}, i.e.,~that one can define the observables in the plane invariant under the translations corresponding to the displacements of the observer point $\mathcal O$ in the plane parallel to the observables. We abandon this latter assumption and leave the position of the observer $\mathcal O$ unchanged. The result we obtain can then be organised so that the leading term in this flat sky approximation explicitly recovers this translational invariance in the plane, as we shall see further below, with the sub-leading terms estimating the measure of the error of such an approximation.

We thus have $\hat \delta(\hat n') \approx \hat \delta(\hat n+\vec \theta)$, and given that the direction $\hat n$ is fixed, we can drop labelling it as a variable, and we simply write $\hat \delta(\hat n') \approx \hat \delta(\vec \theta)$. In a 2D plane, a convenient and often used method for decomposing a 2D field is the Fourier transform
\eeq{
\label{eq:2d_FT}
\hat \df (\vec \ell)  = \int d^2 \theta~ e^{i \vec \ell \cdot \vec \theta} \hat \df (\vec \theta) \, .
}
However, instead of the 2D Fourier transform, we opt for something more in line with the expansion in the spherical harmonics we have used to decompose the spherical shell. The ordinary Bessel functions are a convenient basis for that purpose, given that they are the eigenfunction of the 2D Laplacian in the plane. Thus, representing the 2D position $\vec \theta$ in polar coordinates $\{ \theta, \phi \}$, we obtain
\eeq{
 \hat \df (\vec \theta) = \frac{1}{2\pi} \sum_{m=-\infty}^\infty \int_0^\infty \ell d\ell~ \hat\delta_{m}(\ell) J_m(\ell \theta) e^{- i m \phi}\, ,
}
where the coefficients can be obtained by using the orthogonality of the used eigenfunctions
\eeq{
 \hat\delta_m(\ell) = \int_0^{2\pi} d\phi  \int_0^\infty \theta d\theta ~ J_m(\ell \theta) ~e^{i m \phi}  \hat \df (\vec \theta)\, .
}
This also establishes a natural connection between the flat-sky coefficients $\hat\delta_m(\ell)$ and the full-sky ones $\hat\delta_{\ell,m}$, as the former should approach the latter ones as we approach the field close to the line of sight $\hat n$. 

Using the Jacobi-Anger expansion in Eq.~\eqref{eq:2d_FT} provides us with the connection of the $\hat\delta_m(\ell)$ coefficients and the 2D Fourier transformed field $\hat \df (\vec \ell)$, i.e.,
\eeq{
\hat \df (\vec \ell)  
= \sum_{m=-\infty}^\infty i^m \int d^2 \theta~ J_m(\ell \theta) e^{i m (\phi -\phi_\ell)} \hat \df (\vec \theta)
= \sum_{m=-\infty}^\infty i^m \hat\delta_{m}(\ell) e^{ - i m \phi_\ell} \, .
}
This establishes the correspondence of the two bases in the 2D plane. We can relate these coefficients to the 3D Fourier field 
\eeq{
 \hat\delta_{m}(\ell) = (-i)^{m} \int \frac{d\chi}{\chi^2} ~W(\chi) \int_{-\infty}^\infty \frac{d k_{\hat n}}{2\pi} ~e^{- i \chi  k_{\hat n}} 
  \int_0^{2\pi} \frac{d\phi_k}{2\pi}~ \df \big( k_{\hat n} , \tilde \ell , \phi_k , z[\chi] \big) e^{ i m \phi_k} \, ,
}
where we have separated the dependence of $\df(\vec k)$ field in modes along ($k_{\hat n}$) and perpendicular ($k_{\perp}$) to the line of sight. Modes perpendicular to the line of sight can be additionally decomposed in amplitude $\tilde \ell \equiv k_\perp = \ell/\chi$, and phase $\phi_k$.
We thus have
\eeq{
\hat \df (\vec \ell)  
= \int \frac{d\chi}{\chi^2} ~W(\chi) \int_{-\infty}^\infty \frac{d k_{\hat n}}{2\pi} ~e^{- i \chi  k_{\hat n}} \df \big( k_{\hat n} , \tilde \ell , \phi_\ell , z[\chi] \big)\, ,
}
which is, of course, consistent with the direct 2D Fourier transform in Eq.~\eqref{eq:2d_FT}.

Let us look at the two-point correlator
\eq{
\la \hat\delta_{m}(\ell)  \hat\delta^*_{m'}(\ell')  \ra
&= \df^{K}_{m m'} \int \frac{d\chi_1}{\chi_1^2} \frac{d\chi_2}{\chi_2^2} ~ W(\chi_1)W'(\chi_2) (2\pi) 
\frac{\sqrt{\chi_1\chi_2}}{\sqrt{\ell \ell'}} \df^D \lb \ell/\chi_1 - \ell'/\chi_2 \rb \\
&\hspace{1cm} \times  \int_{-\infty}^\infty \frac{d k_{\hat n}}{2\pi} ~e^{- i ( \chi_1 - \chi_2 )  k_{\hat n}} \mathcal P \lb k_{\hat n}, \sqrt{\ell\ell'}/\sqrt{\chi_1\chi_2}; z[\chi_1], z[\chi_2] \rb\, , \non
}
where we have written the Dirac delta function in polar coordinates as
\eeq{
\df^{\rm 3D} \lb \vec k - \vec k\rb
= \frac{1}{\sqrt{\rho_{k}\rho_{k'}}} \df^{\rm D} \lb \rho_{k} - \rho_{k'} \rb \df^{\rm D} \lb \phi_{k} - \phi_{k'} \rb \df^{\rm D} \lb k_{z} - k'_{z} \rb \, .
}
Moreover, using $\df^{\rm D} \lb \tilde \ell - \tilde \ell' \rb = \sqrt{\chi\chi'} ~ \df^{\rm D}  \big( (\chi' \ell - \chi \ell')/\sqrt{\chi\chi'} \big)$, and thin redshift windows $W(\chi) = \df^D\lb \chi - \chi^* \rb$, we can define the flat-sky version of the \textit{unequal-time angular power spectrum}
\eeq{
\label{eq:C_flat+delta}
\la \hat\delta_{m}(\ell,\chi)  \hat\delta^*_{m'}(\ell',\chi')  \ra =  (2\pi) \frac{\df^{K}_{m m'} }{\sqrt{\ell \ell'}} \mathbb{C}^{\rm flat}\lb \sqrt{\ell\ell'}, \chi, \chi' \rb \df^{\rm D} \lb \frac{\chi' \ell - \chi \ell'}{\sqrt{\chi\chi'}} \rb \, ,
}
with the explicit expression 
\eeq{
\label{eq:C_flat}
\mathbb{C}^{\rm flat} \lb \ell, \chi, \chi' \rb = \frac{1}{\chi \chi'} \int_{-\infty}^\infty \frac{d k_{\hat n}}{2\pi} \, e^{- i \delta \chi  k_{\hat n}} \mathcal P \lb k_{\hat n}, \ell/\sqrt{\chi\chi'}; z[\chi], z[\chi'] \rb\, ,
}
and defining $\delta \chi = \chi - \chi'$. Before we discuss this definition of the flat-sky angular power spectrum $\mathbb{C}^{\rm flat}$ in its possible alternative choices, let us first establish the link between the angular power spectrum obtained using the flat-sky coefficients $\hat\delta_{m}(\ell)$ compared to the 2D Fourier modes $\hat \df (\vec \ell)$. Investigating the correlation function of the 2D Fourier fields $\hat \df (\vec \ell)$ we find 
\eq{
\label{eq:2DFT_vs_bessel}
\la \hat \df(\vec \ell,\chi) \hat \df^*(\vec \ell',\chi') \ra
&= \sum_{m, m'} i^{m-m'} \la \hat\delta_{m}(\ell) \hat\delta^*_{ m'}(\ell') \ra e^{ - i m \phi_{\ell} + i m' \phi_{\ell'}} \\
&=  (2\pi) \mathbb{C}^{\rm flat}\lb \sqrt{\ell\ell'},  \chi, \chi' \rb  \frac{1}{\sqrt{\ell \ell'}} \df^{\rm D} \lb \frac{\chi' \ell - \chi \ell'}{\sqrt{\chi\chi'}} \rb
 \sum_{m} e^{ - i m ( \phi_{\ell} - \phi_{\ell'})} \non\\
&=  (2\pi)^2\mathbb{C}^{\rm flat}\lb \sqrt{\ell\ell'},  \chi, \chi' \rb \df^{\rm 2D} \lb \frac{\chi' \vec{\ell} - \chi \vec{\ell'}}{\sqrt{\chi\chi'}} \rb \, , \non
}
and thus the unequal-time angular power spectrum obtained is equivalent to the one obtained correlating the flat-sky eigenfunction coefficients $\hat\delta_{m}(\ell)$.

Let us comment on the meaning of the Dirac delta function in Eqs.~\eqref{eq:C_flat+delta} and \eqref{eq:2DFT_vs_bessel}. The two modes, $\vec \ell$ and $\vec \ell'$, are modulated by the corresponding comoving distances, keeping in mind that we are comparing the modes on two different redshift slices. These modulations break the translational invariance in the single plane, which would be realised by the simple Dirac delta function of the form $ \df^{\rm D}(\vec \ell - \vec \ell ')$. However, we also know that the corresponding symmetry, i.e.,~statistical isotropy, is realised in the treatment of the entire sky in the form of the Kronecker delta function $\df^{\rm K}_{\ell \ell'}$. We can thus consider the translational invariance of the plane as a flat-sky manifestation of the rotational isotropy of the whole sky, realised in the limiting case of two nearby planes, both distant from the observer. Thus, the deviations from this symmetry that we experience in the flat-sky are a feature of the approximation itself, and the generated off-diagonal terms do not capture any additional full-sky features or contributions. On the contrary, we can interpret the off-diagonal contributions as a measure of the accuracy of our flat-sky approximation, i.e.,~if they are in any way significant or comparable to the diagonal term, we should not expect our flat-sky approximation to be a good representation of the full-sky result.  Furthermore, we expect the result to be physically meaningful when the two planes are close to each other and far from the observer, which motivates us to reorganise our flat-sky result reflecting these properties. Introducing variables $\delta = \delta \chi / (2\bar \chi)$, $\vec \Delta = \vec{\ell}' + \vec{\ell}$ we can write 
\eeq{
\df^{\rm 2D} \lb \frac{\chi' \vec{\ell} - \chi \vec{\ell'}}{\sqrt{\chi\chi'}} \rb = \mathcal A(\delta)\,  \df^{\rm 2D} \lb \vec{\ell} - \vec{\ell'} + \varphi(\delta) \vec \Delta \rb \,,
}
where the obtained amplitude $\mathcal A (\delta)$ and phase $\varphi(\delta)$ are both functions of $\delta$ that depend on a particular choice of the definition of mean comoving distance $\bar \chi$. For a particular choice of the arithmetic mean $\bar \chi_{\rm a}$, we get $\mathcal A = 1-\delta^2$ and $\varphi = \delta$, while with the choice of the harmonic mean $\bar \chi_{\rm h}$ the amplitude shift vanishes, i.e.~$\mathcal A = 1$, at the expense of a more complex phase $\varphi$ dependence on $\delta$.\footnote{For a more detailed discussion of this point we refer the interested reader to the appendix in~\cite{Raccanelli+:2023II}.} Expanding in small $\delta$, i.e.,~around the diagonal contributions, we have 
\eeq{
\df^{\rm 2D} \big( \vec{\ell} - \vec{\ell}' + \varphi(\delta) \vec \Delta  \big)
= \df^{\rm 2D} \big( \vec{\ell} - \vec{\ell}' \big) + 
\Big( e^{\varphi(\delta) \vec \Delta \cdot  \overset{\rightarrow}{\partial}_{\vec \ell}} - 1 \Big) \df^{\rm 2D} \big( \vec{\ell} - \vec{\ell}' \big) \, .
}
The unequal-time two-point correlation function of the projected overdensity field in the Fourier space then becomes 
\eeq{
\label{eq:corr_ell_flat}
\la \hat \df(\vec \ell,\chi) \hat \df^*(\vec \ell',\chi') \ra 
= (2\pi)^2 \df^{\rm 2D} \big( \vec{\ell} - \vec{\ell}' \big) \sum_{n=0}^\infty  \frac{ \big( \overset{\leftarrow}{\partial}_{\vec \ell} \cdot \vec \Delta \big)^n }{n!} 
\mathbb{C}^{(n)} \lb \sqrt{\ell\ell'}, \chi, \chi' \rb\, ,
}
where the $n$-th angular power spectrum is given as
\eeq{
\label{eq:C_n_ell}
\mathbb{C}^{(n)} \lb \ell, \chi, \chi' \rb = \mathcal A(\delta) \phi(\delta)^n\, \mathbb{C}^{\rm flat} \lb \ell, \chi, \chi' \rb \, ,
}
and the expression for the $\mathbb{C}^{\rm flat}$ is given in Eq.~\eqref{eq:C_flat}. Again, if we choose the arithmetic mean $\bar \chi_{\rm a}$, the prefactor  $\mathcal A \phi^n$ simplifies to $(1-\delta^2)\delta^n$, and the higher $n$ terms are thus suppressed by the additional $\delta = \delta \chi / (2\bar \chi)$ terms. Different choices of the mean distance $\bar \chi$ would give somewhat different $\phi$ dependence on $\delta$. Nonetheless, the leading term would still remain linear in $\delta$ (see~\cite{Raccanelli+:2023II}).

Let us consider for a moment the content of this result. As mentioned earlier, Eq.~\eqref{eq:corr_ell_flat} states that in addition to the diagonal term reflecting the translational invariance in the plane, there exist off-diagonal correction terms characterised by the higher derivative operator acting on the Dirac delta function. These correction terms are suppressed with respect to the leading $n=0$ term by the higher powers of $\delta$ (generally $\phi(\delta)$) in Eq.~\eqref{eq:C_n_ell}. The higher $n$ terms arise purely as a consequence of assuming the flat-sky  where the underlying symmetry, namely the isotropy, is realised on a sphere.  As we will discuss below, this suppression not only depends on the weighting of the window functions but is also closely related to the shape of the 3D power spectrum $\mathcal P$ and its wave mode support along the line of sight, which is evident from the integration along the $k_{\hat n}$ in the Eq.~\eqref{eq:C_flat}. The range of support of $\mathcal P$ in $k_{\hat n}$ thus determines the support of $\mathbb{C}^{\rm flat}$ in $\delta$, i.e.,~it determines the extent of unequal-time correlation in the unequal-time angular power spectrum and consequently in the projected angular power spectrum $C(\ell)$. We can estimate the correlation support in the $\delta$ direction by finding the extrema of  $\mathbb{C}^{\rm flat}$, which gives us
\eeq{
\label{eq:condition}
0 = \int_{0}^\infty d k_{\hat n} \,  \sin \lb 2  \delta_* \bar \chi  k_{\hat n} \rb k_{\hat n} \mathcal P \lb k_{\hat n}, \ell/\bar \chi; z[\bar \chi] \rb + \ldots \, 
}
and where we have neglected the other contributions arising from the derivative of the 3D power spectrum and other $\delta$ dependencies. The expression above determines the position of the $\delta_*$ extrema where the unequal-time angular power spectrum exhibits the maximal or minimal correlation strength. We note that this explicitly depends on the shape of the 3D power spectrum. In the next section, we examine these points assuming a simplified shape of the 3D power spectrum. 

Before we close this section, let us further consider the dependence of the unequal-time angular power spectrum on the unequal-time variable $\delta$ and how it affects the integration of the window functions in the projected angular power spectrum. We can also clarify under what conditions the often used Limber approximation~\cite{Limber:54,Kaiser:1991,Kaiser:1996,LoVerde:08} is justified and can be expected to hold. We begin by noting that the unequal-time contributions in the 3D power spectrum can be organised in a series in $\delta$  of the form 
\eeq{
\mathcal P \lb k; z,z' \rb
= \sum_{m=0}^\infty  \mathcal P_m \lb k; \bar z \rb \delta^m \, .
}
This expansion obviously holds in the linear theory $\mathcal P \lb k; z,z' \rb = D(z)D(z') \mathcal P_L(k)$, but also 
in the case of the nonlinear power spectrum, when, e.g.,~higher-order perturbative corrections are considered. Using this expansion in Eq.~\eqref{eq:C_flat} and assuming that $\mathcal P_m$ depend only quadratically on $k_{\hat n}$ (the fact that holds even when redshift-space distortions are taken into account), we get
\eeq{
\mathbb{C}^{\rm flat} \lb \ell, \chi, \chi' \rb 
= \frac{2}{\chi \chi'} \sum_{m=0}^\infty \delta^m
\int_0^\infty \frac{d k_{\hat n}}{2\pi} \, \cos \lb 2 \bar \chi \delta k_{\hat n} \rb
\mathcal P_m \lb k_{\hat n}, \ell/\bar \chi; z[\bar \chi] \rb \, .
}
Integrating over the general window functions $W$ and using the Eq.~\eqref{eq:C_n_ell}, we obtain the projected version of the angular power spectra
\eeq{
C^{(n)}(\ell) = \int d\chi_1 d\chi_2\, W(\chi_1)W'(\chi_2) 
 \mathcal A(\delta) \phi(\delta)^n\, \mathbb{C}^{\rm flat} \lb \ell, \chi_1, \chi_2 \rb \, ,
}
where we define the projected flat-sky angular power spectrum as the leading, diagonal component in our expansion
\eeq{
C(\ell) \equiv C^{(0)}(\ell)  \, .
}
To proceed a bit further, we assume a specific form of the window function, namely, we assume a Gaussian window 
\eeq{
W(\chi) = \frac{1}{\sqrt{2\pi} \sigma} e^{- \frac{(\chi - \chi_*)^2}{2 \sigma^2}} ,
}
which, for the case when both windows in the projected angular power spectrum are equal, gives us 
\eeq{
 W\lb \bar \chi + \tfrac{1}{2} \delta \chi \rb W\lb \bar \chi - \tfrac{1}{2} \delta \chi \rb 
= W\lb \bar \chi \rb^2 
e^{ - \lb \bar \chi/\sigma \rb^2 \delta^2}.
}
In the case when the two windows are not equal, the analysis gets a bit more cumbersome; however, the qualitative results do not change. For the projected angular power spectra, choosing the arithmetic mean, we have
\eq{
\label{eq:C_n_ell_window}
C^{(n)}(\ell) 
&= 2 \int_0^\infty \bar \chi d\bar \chi \, W(\bar \chi)^2 \int_{-\infty}^\infty d\delta\,
 (1-\delta^2)\delta^n e^{ - \lb \bar \chi/\sigma \rb^2 \delta^2} \, \mathbb{C}^{\rm flat} \lb \ell, \bar \chi, \delta \rb  \\
&= 2 \sum_{m=0}^\infty \int_0^\infty \frac{d \bar \chi}{\bar \chi}\, W(\bar \chi)^2  
\int_{-\infty}^\infty \frac{d k_{\hat n}}{2\pi} \, 
G_{n+m} \lb  2 \bar \chi k_{\hat n} , \bar \chi/\sigma  \rb
\mathcal P_m \lb k_{\hat n}, \ell/\bar \chi; z[\bar \chi] \rb \non
  \, ,
}
where we have introduced the kernel $G_N$ containing the integral over $\delta$, that can be written as
\eeq{
G_N (a,b) 
= \int_{-\infty}^\infty d\delta\, \delta^{N} e^{ - i a \delta - b \delta^2} 
= \lb i \partial_a \rb^N G_0(a,b)\, ,
}
and where $G_0(a,b) =  \sqrt{\pi}/b\,  \exp\lb-a^2/(2b)^2\rb$ is the usual Gaussian integral. After a bit of straightforward calculation, we arrive at 
\eeq{
G_N (a,b) = (-i/b)^N U\lb - N/2;\, 1/2;\, a^2/(2b)^2 \rb G_0(a,b)\,,
}
where $U(a;b;z)$ is a confluent hypergeometric function of the second kind.

It is interesting to consider what happens when the window support, characterised by the variance $\sigma$ is large.
The support of the window contributions $G_N$, at large $k_{\hat n}$, is controlled by $G_0$ part that takes the form 
\eeq{
G_{0} \lb  2 \bar \chi k_{\hat n} , \bar \chi/\sigma  \rb =  \sqrt{\pi} \sigma / \bar \chi\,  
\exp\lb- \sigma^2 k_{\hat n}^2\rb \, ,  
}
which effectively constrains the integral domain to $k_{\hat n} \lesssim 1/\sigma$. For some fixed and finite $\ell$ such that $k_{\hat n} \ll \ell/\bar \chi$, one can neglect the $k_{\hat n}$ dependence of the 3D power spectrum. This is, of course, possible since we assume that the 3D power spectrum depends only on the amplitude of the total wave mode $k = \sqrt{k_{\hat n}^2 + \ell^2/\bar \chi^2} \simeq \ell/\bar \chi$, which is not the case when, e.g.,~redshift space distortions are considered (a known regime when Limber approximation fails).  From Eq.~\eqref{eq:C_n_ell_window} we than get 
\eeq{
\label{eq:Cn_ell_Limber}
C^{(n)}(\ell) 
= \sum_{m=0}^\infty  \lb -i \rb^{n+m}  \gamma_{n+m} \int_0^\infty d \bar \chi\, \frac{W(\bar \chi)^2}{\bar \chi^2} 
 \lb \frac{\sigma}{\bar \chi}\rb^{n+m}  \mathcal P_m \lb \ell/\bar \chi; z[\bar \chi] \rb
  \, ,
}
where the integration over the $G_N$ gives us 
\eeq{
\gamma_N = \int_{-\infty}^\infty \frac{d x}{\sqrt{\pi}} \, U\lb -N/2;\, 1/2;\, x^2 \rb  \, e^{- x^2}
= \begin{cases} 
1 & ~~ {\rm if}~ N=0\, ,  \\ 
(-2)^{-N/2+1/2} (N-2)!!/\sqrt{\pi} & ~~ {\rm if}~ N {\rm ~is~odd}\, ,  \\ 
0 & ~~ {\rm if}~ N {\rm ~is~even} \, .
\end{cases} 
}
The usual Limber approximation is obtained by setting $n=0$ and $m=0$, and we obtain the well known expression 
\eeq{
C(\ell)  = C^{(0)}(\ell) =
\int_0^\infty d\bar \chi\, \frac{W(\bar \chi)^2}{\bar \chi^2} \mathcal P \lb \ell/\bar \chi; z[\bar \chi] \rb \, .
}
From the expression in Eq.~\eqref{eq:Cn_ell_Limber} we see that the higher terms in $m$ (and equivalently in $n$), originating from the unequal time effects in the 3D power spectrum, are suppressed in this `large $\sigma$' approximation by the higher powers of $\sigma/\bar \chi$.

In summary, in this section we have established the connection between the full-sky and flat-sky angular power spectrum from a geometrical perspective. Starting from the common propositions of setting the analysis in the plane (flat-sky), we derived the expressions for the angular power spectrum that can be considered a suitable approximation for the full analysis on the sphere (full-sky). The question that naturally arises is under what condition do we expect this approximation to hold? On physical grounds, we expected it to hold for small survey angles (high $\ell$) and distant observers (from the observation planes). In the following sections, we show how these notions arise in a more formal, asymptotic sense and what is the appropriate limit of dynamical variables in which the flat-sky solution is realised. 

\section{Flat-sky limit using the simplified 3D power spectrum}
\label{sec:flat_sky_toy}

In this section, let us focus on the simplified 3D power spectrum. This will allow us to make a straightforward and concrete comparison between the flat-sky and the full-sky results for the unequal-time angular power spectrum $\mathbb C_\ell$. Moreover, by performing the asymptotic expansion around the saddle point, we can establish the precise conditions that give rise to the flat-sky results. Even though obtained in the simplified 3D power spectrum case, these conditions should be independent of the shape and form of the power spectrum and can thus be considered as universal. Indeed, these very conditions are then used in the next section to obtain the flat sky limit for the general case for an arbitrary 3D power spectrum.  

Let us thus assume a form of the 3D power spectrum
\eeq{
\label{eq:toy_PS}
\mathcal P(k; z, z') = A D(z)D(z') k^2 e^{- \alpha^2 k^2}
= \frac{A}{\alpha^2} D(z)D(z')  \lb - \partial_{\kappa} \rb e^{- \kappa k^2} \Big |_{\kappa=\alpha^2} \,.
}
Using this form in the expression given in Eq.~\eqref{eq:C_flat} we obtain a simple and analytic flat-sky unequal-time angular power spectrum 
\eq{
\label{eq:Cell_flat}
\mathbb C^{\rm flat} \lb \ell, \chi, \chi' \rb
 &= A D D' \lb - \partial_{\kappa} \rb e^{ - \kappa  \ell^2/(\chi\chi')}
 \frac{1}{2 \sqrt{\pi \kappa } \chi\chi' } e^{ - \frac{(\delta \chi)^2}{4 \kappa}} \Big |_{\kappa=\alpha^2} \\
 &= \frac{ADD'}{2 \sqrt{\pi} \alpha^3 \chi \chi'}
 \lb \frac{1}{2} + (\alpha \ell)^2/(\chi\chi') - \frac{\df \chi^2}{4 \alpha^2} \rb e^{ - (\alpha \ell)^2/(\chi\chi')  - \frac{\df \chi^2}{4\alpha^2}} \, , \non
} 
and where we use again the notation $\delta \chi = \chi - \chi'$. Using the arithmetic mean $\bar \chi = (\chi+\chi')/2$ and the $\delta$ variable (introduced in the previous section), this expression becomes 
\eeq{
\label{eq:Cell_flat_simple}
\mathbb C^{\rm flat} \lb \ell, \bar \chi, \delta \rb = 
 - \frac{A D D'}{2 \sqrt{\pi} \bar \chi^2(1-\delta^2)}   \partial_{\kappa} \lb
 \frac{1}{\sqrt{\kappa}} \exp \left[  - \frac{\kappa}{\bar \chi^2(1-\delta^2)} \ell^2 - \frac{\bar \chi^2}{\kappa} \delta^2 \right] \rb_{\kappa=\alpha^2} \, .
}

In the full-sky case, the unequal-time angular power spectrum is given by Eq.~\eqref{eq:Cell_full}, which also has an analytic solution for our choice of the 3D power spectrum. We obtain  
\eeq{
\label{eq:Cell_full_simple}
\mathbb{C}^{\rm full}_{\ell} (\bar \chi,\delta) = - \frac{A DD'}{2 \bar \chi \sqrt{1 - \delta^2}} \partial_\kappa \lb \frac{1}{\kappa} 
e^{- \frac{\bar \chi^2}{2\kappa} \lb 1 + \delta^2 \rb} I_{\ell+1/2} \lb \frac{\bar \chi^2}{2 \kappa} \lb 1 - \delta^2 \rb \rb \rb_{\kappa = \alpha^2} \, ,
}
where we use the fact that the integral over the two spherical Bessel functions and Gaussian suppression gives rise to the modified Bessel function $I_\nu(z)$, i.e.
\eeq{
\frac{2}{\pi} \int k^2 d k ~ j_{\ell} (\chi k) j_{\ell} (\chi' k) e^{- \kappa k^2} = \frac{1}{2\sqrt{\chi \chi'}} \frac{1}{\kappa} 
e^{- \frac{\chi^2+\chi'^2}{4 \kappa}} I_{\ell+1/2} \lb \frac{\chi \chi'}{2 \kappa} \rb \, .
}

The question now arises regarding the relation of this full-sky result $\mathbb{C}^{\rm full}_{\ell} $ to the obtained flat-sky result $\mathbb C^{\rm flat} \lb \ell \rb$. 
Our strategy here is to derive the latter from the former. We could, of course, compare them numerically and check the correspondence. However, this is not exactly what we are aiming at, especially since we are dealing with the unrealistic shape of the 3D power spectrum. We want to use the analytic expressions to determine the exact conditions under which the full-sky results approach the flat-sky. This is more valuable information because we will require it to hold universally, regardless of the choice of the shape of our 3D power spectrum. Thus, we are looking for the exact asymptotic limit in which we can recover the flat-sky result $\mathbb C^{\rm flat} \lb \ell \rb$ starting from the expression $\mathbb{C}^{\rm full}_{\ell}$ in Eq.~\eqref{eq:Cell_full_simple}. 

Without further ado, we postulate that the flat-sky results are retrieved in the limit $\chi \propto \ell \to \infty$, i.e.,~when the mean comoving distance $\chi$ is approaching large values as fast as $\ell$ is. We are thus interested in obtaining the asymptotic form of the modified Bessel function $I_\nu(z)$ in that limit, i.e.,~we would like to obtain the approximation for
\eeq{
 I_{\nu+\frac{1}{2}} \lb a \nu^2 - b \rb, ~~{\rm as}~~ \nu \to \infty \, .
}
To do this, we use the standard saddle point method, starting from the integral representation of the modified Bessel function
\eeq{
I_\nu (z)
= \frac{1}{2\pi} \int_{-\pi}^\pi d\theta ~ e^{z \cos \theta + i \nu \theta} \, .
}
We rewrite the integral in the following form
\eeq{
I_{\nu+\frac{1}{2}} \lb a \nu^2 - b \rb 
= \frac{1}{2\pi} \int_{-\pi}^\pi d\theta ~ e^{ i \lb \nu+\frac{1}{2}\rb \theta} e^{ \nu^2 f(\theta)}\, , 
}
where $f(\theta) = \lb a - b/\nu^2 \rb \cos \theta$. From $f'(\theta_0) = - \lb a - b/\nu^2 \rb \sin \theta_0 = 0$ we have $\theta_0 = 0,~\pi,$ and $-\pi$, and thus $f''(\theta_0) = - a + b/\nu^2, ~ a - b/\nu^2$ and $ a - b/\nu^2$. Expanding the integrand around the stationary point $\theta_0 = 0$ gives us
\eeq{
f(\theta) =  \lb a - b/\nu^2 \rb + \frac{1}{2} f''_0 (\theta - \theta_0)^2+\ldots =  \lb a - b/\nu^2 \rb + \frac{1}{2} |f''_0| e^{i \arg (f''_0)} s^2 e^{i 2 \phi} \, ,
}
and we have
\eq{
\label{eq:I_nu_limit}
I_{\nu+\frac{1}{2}} \lb a \nu^2 - b \rb
&\sim \frac{e^{a \nu^2 - b}}{2\pi} \left[ 
\int_{\infty}^0 dt~ e^{ - i \lb \nu+\frac{1}{2}\rb t } e^{- \frac{1}{2} \lb a \nu^2 - b \rb t^2 + \pi}
+ \int_{0}^\infty dt~ e^{ i \lb \nu+\frac{1}{2}\rb t} e^{- \frac{1}{2} \lb a \nu^2 - b \rb t^2}
\right] \non\\
&\sim \frac{ e^{\lb a \nu^2 - b \rb}}{\sqrt{2 \pi \lb a \nu^2 - b \rb}} e^{- \frac{\lb \nu+\frac{1}{2}\rb^2}{2 \lb a \nu^2 - b \rb }}\, , ~~{\rm as}~~ \nu \to \infty\, .
}
We have thus obtained the leading asymptotic term by expanding around the stationary $\theta=0$ point. Further corrections could be obtained by exploring the subleading terms of this saddle, as well as by considering the corrections arising from the borders of the integration region. 

Using $\nu=\ell$, $a= \frac{\bar \chi^2}{2 \kappa \ell^2}$ and $b= \frac{\bar \chi^2}{2\kappa} \delta^2$, we obtain 
\eeq{
I_{\ell+1/2} \lb \frac{\bar \chi^2}{2 \kappa} \lb 1 - \delta^2 \rb \rb 
\approx \frac{\sqrt{\kappa}}{\bar \chi}
 \frac{ e^{ \frac{\bar \chi^2}{2 \kappa} \lb 1 - \delta^2 \rb}}{\sqrt{\pi \lb 1 - \delta^2 \rb}} \exp\left[ - \frac{\kappa}{\bar \chi^2 (1 - \delta^2)}\ell'^2 \right] \, ,
}
where we use $\ell' = \ell + 1/2$, which also gives us a mathematical justification of the commonly used approximation (often adopted in conjunction with the Limber approximation, see, e.g.,~\cite{LoVerde:08}). This gives us the limit for the full-sky unequal-time angular power spectrum
\eq{
\mathbb{C}^{\rm full}_{\ell}  (\bar \chi, \delta) \approx - \frac{A DD'}{2 \sqrt{\pi} \bar \chi^2 (1 - \delta^2)} \partial_\kappa \lb \frac{1}{\sqrt \kappa} 
\exp\left[- \frac{\bar \chi^2}{\kappa} \delta^2 - \frac{\kappa}{\bar \chi^2 (1 - \delta^2)}\ell'^2\right]
 \rb_{\kappa = \alpha^2} \, ,
}
which is equivalent to the flat-sky result $\mathbb C^{\rm flat} \lb \ell \rb$ given in Eq.~\eqref{eq:Cell_flat_simple}, up to the difference in $\ell$ and $\ell'$. This thus justifies our limiting procedure where we assumed $\chi \propto \ell \to \infty$. In the next section, we show how we can generalise these results to the case of a general 3D power spectrum. Luckily, it turns out we have done most of the calculations that we will need already in this section.

Before we move on, however, let us use our simple example to investigate the support of the unequal-time effects in the flat-sky angular power spectrum $\mathbb C^{\rm flat} \lb \ell \rb$. Condition for finding the extrema is given in Eq.~\eqref{eq:condition}, which, besides the trivial $\delta_*= 0$ solution, gives us two finite $\delta_*$ solutions
\eeq{
\delta_* \approx \pm \frac{\alpha}{\bar \chi} \sqrt{ \frac{3}{2}\lb 1 + \frac{2 (\alpha \ell)^2}{3 \bar \chi^2} \rb} \, .
}
The mathematical details of this expression are not highly important by themself; nonetheless, there are several lessons to be learned. First of all, the flat-sky angular power spectrum $\mathbb C^{\rm flat} \lb \ell \rb$ at these two points is negative, i.e.,~comparing the two specific time slices (separated approximately by $\delta_*$) structure is anti-correlated  (see also~\cite{Raccanelli+:2023I}). What determines this anti-correlation length? We see that $\delta_*$ is dependent on $\alpha$, i.e.,~it is determined by the shape of the 3D power spectrum $\mathcal P$. We thus expect to find similar features in more general, $\Lambda$CDM-like, cosmologies.

\section{Flat-sky limit using the general 3D power spectrum}
\label{sec:flat_sky_limit_general}

We generalise our results from the previous section, obtained by considering a simplified shape of the 3D power spectrum, to the case of the general shape, which includes the realistic $\Lambda$CDM power spectrum. We start with the assumption that the 3D power spectrum $\mathcal P$ can be represented as a discrete Mellin transform of the following form
\eeq{
\label{eq:Pk_Mellin}
\mathcal P(k;\, \chi,\chi') = D D' \sum_{i} \alpha_i k^{\nu_i} \, ,
}
where $\alpha_i$ are coefficients, and the $\nu_i$ are phases. Each of these can be complex. This is a good approximation that works well in various LSS applications (see, e.g.~\cite{Hamilton:1999, Schmittfull+:2016_I, Schmittfull+:2016_II, McEwen++:2016, Assassi++:2017, Simonovic++:2017} for some examples). Our strategy is thus to use this transform and perform the limiting procedure on the individual $k^{\nu_i}$ case. 

Adopting the transform given in Eq.~\eqref{eq:Pk_Mellin} and using it in the flat-sky angular power spectrum expression given in Eq.~\eqref{eq:C_flat}, we get
\eeq{ 
\label{eq:C_flat_K}
\mathbb C^{\rm flat}(\ell, \chi, \chi')
 = \frac{D D'}{\chi\chi'} \sum_i \alpha_i
 \int_{-\infty}^\infty \frac{d k_{\hat n}}{2\pi} ~ e^{-i \delta \chi k_{\hat n}}
\lb k_{\hat n}^2+{\tilde\ell}^2 \rb ^{\frac{\nu_i}{2}}
= \frac{D D'}{\chi\chi'} \sum_i \alpha_i\, \frac{(2\tilde{\ell}/|\delta\chi|)^{\frac{\nu_i}{2}+\frac{1}{2}}}{\sqrt{\pi}\Gamma(-\frac{\nu_i}{2})} 
K_{\frac{\nu_i}{2}+\frac{1}{2}}(|\delta\chi|\tilde\ell)\, ,
}
where we use $\tilde\ell = \ell/\sqrt{\chi \chi'}$, and where $K_\nu$ is the modified Bessel function of the second kind. This expression is useful because it allows efficient evaluation of the calculation of the projected angular power spectrum. It is analogous to the computation performed in reference~\cite{Assassi++:2017} for the full-sky case, where the ordinary hypergeometric function ${ }_2F_1$ is obtained instead of $K_\nu$. In this respect, the flat-sky results provide us with significant computational simplification. However, we won't discuss these aspects of the flat-sky results here; for a detailed analysis and performance of these flat-sky results, we refer the reader to reference~\cite{Gao++:2023}. Here we attempt to obtain the result in Eq.~\eqref{eq:C_flat_K} directly from the full-sky formalism.

Referring to the full-sky unequal-time angular power spectrum expression given in Eq.~\eqref{eq:Cell_full} and using the discrete transform of the 3D power spectrum given in Eq.~\eqref{eq:Pk_Mellin}, we have
\eq{
\label{eq:C_ell_full_Mellin}
\mathbb{C}^{\rm full}_{\ell}  (\chi,\chi') = 4\pi D D' \sum_{i} \alpha_i 
\int_0^\infty \frac{k^2 dk}{2\pi^2}\;  k^{\nu_i}\, j_\ell(k \chi) j_\ell(k \chi') \, .
}
We can use the integral representation of the product of two spherical Bessel functions
\eeq{
j_{\nu}(z) j_{\nu}(\zeta)  = \frac{1}{4 i \sqrt{z \zeta}}
\int_{c-i\infty}^{c+i\infty} \frac{\,\mathrm{d}t}{t} \*\exp \lb \frac{1}{2}t-\frac{z^{2}+\zeta^{2}}{2t} \rb 
I_{\nu+1/2} \lb \frac{z\zeta}{t} \rb \, ,~~~~\Re(\nu)>-1/2\, ,
}
where $c$ is a positive constant (see, e.g.~\cite[\href{http://dlmf.nist.gov/10.9.E28}{(10.9.28)}]{NIST:DLMF}). Using our earlier result on the asymptotic expansion of the modified Bessel function given in Eq.~\eqref{eq:I_nu_limit}, we have
\eeq{
j_{\nu}(z) j_{\nu}(\zeta) 
\approx - \frac{1}{4} \frac{i}{\sqrt{2\pi}} \frac{1}{z\zeta} 
\int_{c-i\infty}^{c+i\infty} \frac{\,\mathrm{d}t}{\sqrt{t}} \*\exp \lb \frac{1}{2}t-\frac{(z-\zeta)^2}{2t} \rb 
  e^{- \frac{\nu'^2}{2 z\zeta} t } \, .
}
This gives us
\eq{
4\pi \int_0^\infty  \frac{k^2 dk}{2\pi^2}\;  k^{\nu}\, j_\ell(k \chi) j_\ell(k \chi') 
&\approx - i \sqrt{\frac{\pi}{2}} \frac{1}{\chi\chi'} 
\int_0^\infty  \frac{dk}{2\pi^2}\;  k^{\nu}\,\int_{c-i\infty}^{c+i\infty}\*\exp \lb \frac{1}{2}t-\frac{\delta \chi^2}{2t} k^2 \rb 
  e^{- \frac{\ell'^2}{2 \chi\chi' k^2 }  t }\
 \frac{\,\mathrm{d}t}{\sqrt{t}}\, \\
&\approx - i \sqrt{\frac{\pi}{2}} \frac{1}{\chi\chi'} 
\int_{c-i\infty}^{c+i\infty} \frac{\,\mathrm{d}t}{\sqrt{t}}\, e^{t/2} 
 \int_0^\infty  \frac{dk}{2\pi^2}\;  k^{\nu}
 e^{ - \frac{\delta \chi^2}{2t} k^2 - \frac{\ell'^2}{2 \chi\chi' k^2 }  t } \, \non\\
&\approx - \frac{i}{(2 \pi)^{3/2}}\frac{1}{\chi\chi'} 
\lb \tilde \ell' / |\delta \chi| \rb^{\frac{\nu}{2} + \frac{1}{2}} K_{\frac{\nu}{2} + \frac{1}{2}} \lb |\delta \chi| \tilde \ell' \rb
\int_{c-i\infty}^{c+i\infty} \mathrm{d}t \, e^{t/2} t^{\frac{\nu}{2}} \, , \non
}
where $\tilde \ell' = (\ell + 1/2)/\sqrt{\chi\chi'}$, and in the second line, we have used the integral representation of the modified Bessel function of the second kind (see, e.g.,~\cite[\href{http://dlmf.nist.gov/10.32.E10}{(10.32.10)}]{NIST:DLMF}. Finally, the remaining integral is related to the definition of the gamma function
\eeq{
\frac{1}{2\pi i} \int_{c-i\infty}^{c+i\infty} \mathrm{d}t \, e^{t/2} t^{s} = \frac{2^{s+1}}{\Gamma(-s)} \, .
}
This gives us an asymptotic form of the full-sky unequal-time angular power spectrum 
\eeq{
\label{eq:C_ell_asym}
\mathbb{C}^{\rm full}_{\ell}  (\chi,\chi') = 4\pi D D' \sum_{i} \alpha_i \int_0^\infty  \frac{k^2 dk}{2\pi^2}\;  k^{\nu}\, j_\ell(k \chi) j_\ell(k \chi') 
\approx \frac{D D' }{\chi\chi'}\sum_{i} \alpha_i  \frac{\lb 2 \tilde \ell' / |\delta \chi| \rb^{\frac{\nu_i}{2} + \frac{1}{2}} }{\sqrt{\pi} \Gamma\lb -\frac{\nu_i}{2}\rb} 
K_{\frac{\nu_i}{2} + \frac{1}{2}} \lb |\delta \chi| \tilde \ell' \rb \, .
}
This is precisely the same form we have obtained from the flat-sky calculations in Eq.~\eqref{eq:C_flat_K}. 

Thus we have established a direct mathematical correspondence between the full- and the flat-sky angular power spectrum. We have already shown that this is to be expected on physical grounds in Sec. \ref{sec:unequal_time}, where geometric considerations are used to establish the correspondence between the full- and the flat-sky for distant observers looking at the small patch of the sky. Here we have succeeded in showing that the same correspondence follows purely mathematically when the proper limit of large $\chi$ and $\ell$ variables is taken, namely when $\chi \propto \chi' \propto \ell \to \infty$, as we have shown in Sec.~\eqref{sec:flat_sky_toy}.

\section{Asymptotic form of the ordinary hypergeometric function ${}_2F_1(a, b ; c; z)$}
\label{sec:math_result}

Here we establish a link between a direct representation of the double spherical Bessel integral given in Eq.~\eqref{eq:C_ell_full_Mellin} given in terms of the ordinary (Gaussian) hypergeometric function ${}_2F_1(a, b ; c; z)$ and, on the other hand, our asymptotic representation established in the Sec.~\ref{sec:flat_sky_limit_general}. This gives us an asymptotic expansion of the ordinary hypergeometric function ${}_2F_1$ in a distinct variable regime that we further specify in this section. We show how our asymptotic result can also be obtained in an alternative way using some of the known properties of the ordinary hypergeometric function.

We start with the representation of the double spherical Bessel integral arising in Eq.~\eqref{eq:C_ell_full_Mellin} in terms of the ordinary hypergeometric function, giving us 
\eeq{
\label{eq:double_bessel_2F1}
 4\pi \int_0^\infty \frac{k^2 dk}{2\pi^2}\, k^{\nu} j_\ell(k \chi) j_\ell(k \chi')
= 2^{\nu+1} \chi^{-3-\nu} \frac{\Gamma\lb \ell +\frac{\nu}{2} + \frac{3}{2} \rb}{\Gamma\lb - \frac{\nu}{2} \rb  \Gamma\lb \ell + \frac{3}{2} \rb } 
t^\ell {}_2F_1 \lb \tfrac{\nu}{2} + 1 , \ell + \tfrac{\nu}{2} + \tfrac{3}{2} ; \ell+\tfrac{3}{2}; t^2 \rb, ~~~{\rm for}~~~{t \leq 1} \, ,
}
where $t= \chi'/\chi = (1-\delta)/(1+\delta)$, and we assume without loss of generality that $\chi \geq \chi'$ (positive $\delta$). This explicit result has already been used to compute the CMB and LSS angular statistics~\cite{Assassi++:2017, Grasshorn++:2017, Schoneberg++:2018, Chen++:2021}. However, the complexity of the hypergeometric function poses limits to the efficiency of using this result. As shown in~\cite{Gao++:2023}, for all practical purposes in CMB and LSS, replacing the full result with its asymptotic form, as given in the previous section, leads to highly accurate results while significantly reducing the computational effort. The reason for this simplification lies in replacing the hypergeometric function with the modified Bessel function of the second kind.

Given our asymptotic results provided in Eq.~\eqref{eq:C_ell_asym} and the analytic expression in Eq.~\eqref{eq:double_bessel_2F1}, we can establish the following relation
\eeq{
\label{eq:2F1_asym}
{}_2F_1 \lb \lambda , \ell + \lambda ; \ell+1; t^2 \rb
\sim \frac{1}{\sqrt{\pi}}
 \lb \frac{\ell}{2} \rb^{\lambda - \frac{1}{2}}  \frac{\Gamma\lb \ell + 1 \rb } {\Gamma\lb \ell + \lambda \rb}
t^{-\ell - \frac{\lambda}{2} - \frac{1}{4}} \lb 1-t \rb^{-\lambda + \frac{1}{2}}
K_{\lambda - \frac{1}{2}} \left[ \ell (1-t)/\sqrt{t} \right] \,, ~~{\rm for} ~~ \ell \to \infty\,, 
}
and for $t\leq1$. However, in order to make these results fully consistent, we need to impose certain conditions on the variable $t$. Namely, when deriving the saddle point approximation for the modified Bessel function $I_\nu$ in Sec.~\ref{sec:flat_sky_toy}, we imposed the condition that $\chi \propto \chi' \propto \ell$, i.e.,~that the variable measuring the magnitude of unequal-time effects, $\delta = \delta \chi/(2 \bar \chi) \propto 1/\ell$, is small. This implies that the above results in Eq.~\eqref{eq:2F1_asym} are valid for small values of $t$. How small? We can relate the smallness to the $\ell$ variable, $\delta = \delta \chi/(2 \bar \chi) =  \delta \chi k_\perp/(2 \ell) = x/(2\ell)$, where we have introduced an arbitrary constant $x(\equiv\delta \chi k_\perp)$. Thus, the above results hold for any $x$ such that $x \ll \ell$. Therefore, the asymptotic result in Eq.~\eqref{eq:2F1_asym} holds when 
\eeq{
t = \frac{1 - x/(2\ell)}{1 + x /(2\ell)}\, ,
}
for an arbitrary $x$. We can simplify the above results further by noting that we are working in the $\ell \to \infty$ limit, and thus $t\sim 1 - x/\ell$. The Eq.~\eqref{eq:2F1_asym} then simplifies to the following compact form 
\eeq{
\label{eq:2F1_asym_final}
{}_2F_1 \lb \lambda , \ell + \lambda ; \ell+1; 1 - \frac{2x}{\ell} \rb
\sim \sqrt{\frac{2}{\pi}}
\lb \frac{\ell}{2}\rb^\lambda \lb 1 + \frac{x}{\ell} \rb^{\ell}
\frac{K_{\lambda - \frac{1}{2}} \big(x\big)}{ x^{\lambda - \frac{1}{2}} }  \,, ~~{\rm for} ~~ \ell \to \infty\, .
}
This result represents the asymptotic expansion of the ordinary hypergeometric function in a specific configuration defined above. Moreover, in the strict limit, when $\ell \gg x$, we can further simplify this result using $\lb 1 + x/\ell \rb^{\ell} \to \exp(x)$. Further asymptotic relations, valid in different variable domains, can be obtained using various transformation rules of the ordinary hypergeometric function. The most immediate ones follow from applying, e.g.,~Euler and Pfaff transformations. 

Moreover, with the hindsight of our previous result, we can also derive the asymptotic expression given in Eq.~\eqref{eq:2F1_asym_final} using some of the well-known relations for the ordinary hypergeometric function ${ }_2F_1$. The essential piece of information is to understand the appropriate limit to be considered, which of course, follows from the insight provided in Sec.~\ref{sec:flat_sky_toy}. Thus, we can use the linear transformation property
\eq{
{}_2F_1 \lb a, b; c; z \rb
=(1-z)^{-a}  {}_2F_1 \lb  a, c-b \;  c ; z/(z-1) \rb \, , 
}
and the relation to Tricomi's (confluent hypergeometric) function $U(a,b,z)$ giving us
\eeq{
U(a,b,z) = z^{-a} \lb \lim_{c\to\infty}  {}_2F_1 \lb a, a - b + 1; c; 1 - c/z \rb \rb \, .
}
The latter relation is sometimes also used as a definition of the function $U(a,b,z)$. Combining the two relations above gives us
\eeq{
{}_2F_1 \lb \lambda , \ell + \lambda ;  \ell+1 ; 1 - 2x/\ell \rb
=\lb \frac{2x}{\ell}\rb^{-\lambda} {}_2F_1 \lb \lambda, 1 - \lambda;  \ell+1; 1 - \ell/(2x)\right)
\sim \ell^{\lambda}\, U \lb \lambda, 2\lambda ; 2x \rb \, , ~~{\rm for}~~ \ell \to \infty \, .
}
The final step is to note that for $b = 2a$, the Tricomi's (confluent hypergeometric) function can be related to the modified Bessel function of the second kind 
\eeq{
U\left(\lambda,2\lambda,2z\right)=\frac{1}{\sqrt{\pi}}e^{z}\lb 2z \rb^{-\lambda + 1/2}K_{\lambda - 1/2} \lb z \rb \, .
}
Combining all these parts gives us 
\eq{
\label{eq:2F1_asym_final_II}
{}_2F_1 \lb \lambda , \ell + \lambda ;  \ell+1 ; 1 - 2x/\ell \rb
&\sim  \sqrt{\frac{2}{\pi}} \lb \frac{\ell}{2} \rb^{\lambda} \frac{ e^x\,  K_{\lambda - 1/2}\lb x \rb}{x^{\lambda - 1/2}} \, , ~~{\rm for}~~ \ell \to \infty \, ,
}
which is equivalent to the result obtained in Eq.~\eqref{eq:2F1_asym_final}.

It is easy to verify that starting from the expression given in Eq.~\eqref{eq:2F1_asym_final} (and equivalently in Eq.~\eqref{eq:2F1_asym_final_II}) and using it in Eq.~\eqref{eq:double_bessel_2F1}, we can recover our flat-sky result from the previous section given in Eq.~\eqref{eq:C_ell_asym}.

\section{Summary and Conclusions} 
\label{sec:conclusion}

We have established a robust link between the full- and flat-sky descriptions of the leading two-point statistics used in CMB and LSS data analyses, namely the angular power spectrum. So far, the two main modes of employing the angular power spectrum have been within the Limber approximation or the so-called full-sky implementation, the latter requiring a costly evaluation of oscillatory integrals containing a product of two Bessel functions.

The Limber approximation is foremost a practical approach. However, it has two considerable drawbacks. The first is related to the fact that it yields results with the required accuracy only at relatively small scales and for surveys with fairly wide windows. Moreover, the accuracy is  limited to auto-correlations. The reason for these limitations is the assumption that the wave modes along the line-of-sight (crucial in the RSDs) in the 3D power spectrum can be entirely neglected when computing the angular power spectrum. Given these issues, the use of the Limber approximation is restricted mainly to the analysis of weak gravitational galaxy and CMB lensing. At the same time, they represent a serious obstacle to effective cosmological analyses using galaxy tomography. On the other hand, the full-sky results are exact and capture the whole angular dependence of the correlators. However, the evaluation of integrals involving a product of two Bessel functions and a theoretical 3D power spectrum has proven to be a challenging task, especially when part of MCMC analyses. In recent years, there have been advances proposing the use of the discrete Mellin transform (known in the field as the FFTLog decomposition), which leads to an expression in terms of the ordinary hypergeometric function ${}_2F_1$ (see notably~\cite{Assassi++:2017}). Although this representation is exact, it still poses a computational challenge when used at all scales, especially in the regime of high $\ell$ or for distant observers (complementary to the Limber approximation). To mitigate this, one might consider patching the full-sky result with the Limber approximation to achieve efficient yet satisfactory accuracy on overall scales. However, this quickly becomes a sensitive fine-tuning problem, with challenges increasing for unequal-time cross-correlations and narrow windows.

Given this state of affairs, we found it prudent to develop a systematic framework that mitigates these challenges and provides a robust and natural way to connect large- and small-scale results. This is achieved by providing an asymptotic approximation to the full-sky result in the limiting case of a distant observer and a large $\ell$ expansion that is consistent with the unequal-time results obtained in the flat-sky limit, i.e.,~when the analysis on spherical shells is replaced by the parallel planes. To establish the precise asymptotic conditions under which this limit is achieved, we first consider a simple analytic form of the theoretical 3D power spectrum. This allows us to obtain the analytic expressions for both the full- and flat-sky angular power spectra, which in turn gives a limiting procedure that maps the former to the latter. It turns out that the full-sky result matches the flat-sky result (in the leading saddle-point approximation) when the angular modes $\ell$ and comoving distance are taken to infinity at the same rate, i.e.,~when $\chi \propto \chi' \propto \ell \to \infty$. The final step is to use this well-defined asymptotic limit to derive the expansion of the full-sky angular power spectrum valid for the general theoretical 3D power spectra (which would also include $\Lambda$CDM-like universes). The latter is automatically achieved by using the above-mentioned Mellin integral transform of the theoretical 3D power spectrum. As a result, we find that the same limiting procedure of high $\ell$ and distant observer naturally maps the full-sky result into the one obtained in the flat-sky approximation, which establishes a robust asymptotic connection between the two results, allowing us also to consider subleading asymptotic corrections, a task we leave for future work.

By deriving our asymptotic connection between the full- and flat-sky angular power spectra, we have also derived a purely mathematical result. Namely, our analysis establishes an asymptotic limit of the ordinary hypergeometric function in the specific variable configuration (corresponding to the flat-sky limit in our physical interpretation). In this configuration, the limit connects the hypergeometric function ${}_2F_1(a,b;c;z)$ to the modified Bessel function of the second kind $K_\nu(z)$. Once the nature of the appropriate limit is established, namely that $\chi \propto \chi' \propto \ell \to \infty$, we are able to provide an alternative derivation of our earlier flat-sky result using some of the known properties and limits of the ordinary hypergeometric function.

\begin{acknowledgments}
We would like to thank Anthony Challinor for the useful discussions and comments. AR acknowledges funding from the Italian Ministry of University and Research (MIUR) through the ``Dipartimenti di eccellenza'' project ``Science of the Universe''. ZV is partially supported by the Kavli Foundation.
\end{acknowledgments}

\section*{References}
\bibliography{ms} 

\begin{thebibliography}{33}
\expandafter\ifx\csname natexlab\endcsname\relax\def\natexlab#1{#1}\fi
\expandafter\ifx\csname bibnamefont\endcsname\relax
  \def\bibnamefont#1{#1}\fi
\expandafter\ifx\csname bibfnamefont\endcsname\relax
  \def\bibfnamefont#1{#1}\fi
\expandafter\ifx\csname citenamefont\endcsname\relax
  \def\citenamefont#1{#1}\fi
\expandafter\ifx\csname url\endcsname\relax
  \def\url#1{\texttt{#1}}\fi
\expandafter\ifx\csname urlprefix\endcsname\relax\def\urlprefix{URL }\fi
\providecommand{\bibinfo}[2]{#2}
\providecommand{\eprint}[2][]{\url{#2}}

\bibitem[{\citenamefont{Aghanim et~al.}(2020)}]{Planck++:2018}
\bibinfo{author}{\bibfnamefont{N.}~\bibnamefont{Aghanim}} \bibnamefont{et~al.}
  (\bibinfo{collaboration}{Planck}), \bibinfo{journal}{Astron. Astrophys.}
  \textbf{\bibinfo{volume}{641}}, \bibinfo{pages}{A1} (\bibinfo{year}{2020}),
  \eprint{1807.06205}.

\bibitem[{\citenamefont{Limber}(1954)}]{Limber:54}
\bibinfo{author}{\bibfnamefont{D.~N.} \bibnamefont{Limber}},
  \bibinfo{journal}{Astrophys. J.} \textbf{\bibinfo{volume}{119}},
  \bibinfo{pages}{655} (\bibinfo{year}{1954}).

\bibitem[{\citenamefont{Kaiser}(1984)}]{Kaiser:1984}
\bibinfo{author}{\bibfnamefont{N.}~\bibnamefont{Kaiser}},
  \bibinfo{journal}{Astrophys. J. Lett.} \textbf{\bibinfo{volume}{284}},
  \bibinfo{pages}{L9} (\bibinfo{year}{1984}).

\bibitem[{\citenamefont{Laureijs et~al.}(2011)}]{Euclid:2011}
\bibinfo{author}{\bibfnamefont{R.}~\bibnamefont{Laureijs}} \bibnamefont{et~al.}
  (\bibinfo{collaboration}{EUCLID}) (\bibinfo{year}{2011}), \eprint{1110.3193}.

\bibitem[{\citenamefont{Aghamousa et~al.}(2016)}]{DESI:2016}
\bibinfo{author}{\bibfnamefont{A.}~\bibnamefont{Aghamousa}}
  \bibnamefont{et~al.} (\bibinfo{collaboration}{DESI}) (\bibinfo{year}{2016}),
  \eprint{1611.00036}.

\bibitem[{\citenamefont{Dor\'e et~al.}(2014)}]{Dore:2014}
\bibinfo{author}{\bibfnamefont{O.}~\bibnamefont{Dor\'e}} \bibnamefont{et~al.}
  (\bibinfo{year}{2014}), \eprint{1412.4872}.

\bibitem[{\citenamefont{Abdalla et~al.}(2015)}]{Abdalla:2015}
\bibinfo{author}{\bibfnamefont{F.~B.} \bibnamefont{Abdalla}}
  \bibnamefont{et~al.} (\bibinfo{collaboration}{Cosmology SWG})
  (\bibinfo{year}{2015}), \eprint{1501.04035}.

\bibitem[{\citenamefont{Abell et~al.}(2009)}]{Abell:2009}
\bibinfo{author}{\bibfnamefont{P.~A.} \bibnamefont{Abell}} \bibnamefont{et~al.}
  (\bibinfo{collaboration}{LSST Science, LSST Project}) (\bibinfo{year}{2009}),
  \eprint{0912.0201}.

\bibitem[{\citenamefont{Leonard et~al.}(2023)}]{Leonard:2022}
\bibinfo{author}{\bibfnamefont{C.~D.} \bibnamefont{Leonard}}
  \bibnamefont{et~al.} (\bibinfo{collaboration}{LSST Dark Energy Science}),
  \bibinfo{journal}{Open Journal of Astrophysics} \textbf{\bibinfo{volume}{6}},
  \bibinfo{pages}{1} (\bibinfo{year}{2023}), \eprint{2212.04291}.

\bibitem[{\citenamefont{Raccanelli and
  Vlah}(2023{\natexlab{a}})}]{Raccanelli+:2023I}
\bibinfo{author}{\bibfnamefont{A.}~\bibnamefont{Raccanelli}} \bibnamefont{and}
  \bibinfo{author}{\bibfnamefont{Z.}~\bibnamefont{Vlah}}
  (\bibinfo{year}{2023}{\natexlab{a}}), \eprint{2305.16278}.

\bibitem[{\citenamefont{Raccanelli and
  Vlah}(2023{\natexlab{b}})}]{Raccanelli+:2023II}
\bibinfo{author}{\bibfnamefont{A.}~\bibnamefont{Raccanelli}} \bibnamefont{and}
  \bibinfo{author}{\bibfnamefont{Z.}~\bibnamefont{Vlah}}
  (\bibinfo{year}{2023}{\natexlab{b}}), \eprint{2306.00808}.

\bibitem[{\citenamefont{Seljak}(1997)}]{Seljak:1996}
\bibinfo{author}{\bibfnamefont{U.}~\bibnamefont{Seljak}},
  \bibinfo{journal}{Astrophys. J.} \textbf{\bibinfo{volume}{482}},
  \bibinfo{pages}{6} (\bibinfo{year}{1997}), \eprint{astro-ph/9608131}.

\bibitem[{\citenamefont{Zaldarriaga and Seljak}(1997)}]{Zaldarriaga+:1996}
\bibinfo{author}{\bibfnamefont{M.}~\bibnamefont{Zaldarriaga}} \bibnamefont{and}
  \bibinfo{author}{\bibfnamefont{U.}~\bibnamefont{Seljak}},
  \bibinfo{journal}{Phys. Rev. D} \textbf{\bibinfo{volume}{55}},
  \bibinfo{pages}{1830} (\bibinfo{year}{1997}), \eprint{astro-ph/9609170}.

\bibitem[{\citenamefont{Kamionkowski et~al.}(1997)\citenamefont{Kamionkowski,
  Kosowsky, and Stebbins}}]{Kamionkowski++:1996}
\bibinfo{author}{\bibfnamefont{M.}~\bibnamefont{Kamionkowski}},
  \bibinfo{author}{\bibfnamefont{A.}~\bibnamefont{Kosowsky}}, \bibnamefont{and}
  \bibinfo{author}{\bibfnamefont{A.}~\bibnamefont{Stebbins}},
  \bibinfo{journal}{Phys. Rev. D} \textbf{\bibinfo{volume}{55}},
  \bibinfo{pages}{7368} (\bibinfo{year}{1997}), \eprint{astro-ph/9611125}.

\bibitem[{\citenamefont{Hu and White}(1997)}]{Hu+:1997}
\bibinfo{author}{\bibfnamefont{W.}~\bibnamefont{Hu}} \bibnamefont{and}
  \bibinfo{author}{\bibfnamefont{M.~J.} \bibnamefont{White}},
  \bibinfo{journal}{Phys. Rev. D} \textbf{\bibinfo{volume}{56}},
  \bibinfo{pages}{596} (\bibinfo{year}{1997}), \eprint{astro-ph/9702170}.

\bibitem[{\citenamefont{Hu}(2000)}]{Hu:2000}
\bibinfo{author}{\bibfnamefont{W.}~\bibnamefont{Hu}}, \bibinfo{journal}{Phys.
  Rev. D} \textbf{\bibinfo{volume}{62}}, \bibinfo{pages}{043007}
  (\bibinfo{year}{2000}), \eprint{astro-ph/0001303}.

\bibitem[{\citenamefont{Lewis and Challinor}(2006)}]{Lewis+:2006}
\bibinfo{author}{\bibfnamefont{A.}~\bibnamefont{Lewis}} \bibnamefont{and}
  \bibinfo{author}{\bibfnamefont{A.}~\bibnamefont{Challinor}},
  \bibinfo{journal}{Phys. Rept.} \textbf{\bibinfo{volume}{429}},
  \bibinfo{pages}{1} (\bibinfo{year}{2006}), \eprint{astro-ph/0601594}.

\bibitem[{\citenamefont{Assassi et~al.}(2017)\citenamefont{Assassi, Simonović,
  and Zaldarriaga}}]{Assassi++:2017}
\bibinfo{author}{\bibfnamefont{V.}~\bibnamefont{Assassi}},
  \bibinfo{author}{\bibfnamefont{M.}~\bibnamefont{Simonović}},
  \bibnamefont{and}
  \bibinfo{author}{\bibfnamefont{M.}~\bibnamefont{Zaldarriaga}},
  \bibinfo{journal}{Journal of Cosmology and Astroparticle Physics}
  \textbf{\bibinfo{volume}{2017}}, \bibinfo{pages}{054–054}
  (\bibinfo{year}{2017}), ISSN \bibinfo{issn}{1475-7516},
  \urlprefix\url{http://dx.doi.org/10.1088/1475-7516/2017/11/054}.

\bibitem[{\citenamefont{Sch\"oneberg et~al.}(2018)\citenamefont{Sch\"oneberg,
  Simonovi\'c, Lesgourgues, and Zaldarriaga}}]{Schoneberg++:2018}
\bibinfo{author}{\bibfnamefont{N.}~\bibnamefont{Sch\"oneberg}},
  \bibinfo{author}{\bibfnamefont{M.}~\bibnamefont{Simonovi\'c}},
  \bibinfo{author}{\bibfnamefont{J.}~\bibnamefont{Lesgourgues}},
  \bibnamefont{and}
  \bibinfo{author}{\bibfnamefont{M.}~\bibnamefont{Zaldarriaga}},
  \bibinfo{journal}{JCAP} \textbf{\bibinfo{volume}{10}}, \bibinfo{pages}{047}
  (\bibinfo{year}{2018}), \eprint{1807.09540}.

\bibitem[{\citenamefont{Grasshorn~Gebhardt and Jeong}(2018)}]{Grasshorn++:2017}
\bibinfo{author}{\bibfnamefont{H.~S.} \bibnamefont{Grasshorn~Gebhardt}}
  \bibnamefont{and} \bibinfo{author}{\bibfnamefont{D.}~\bibnamefont{Jeong}},
  \bibinfo{journal}{Phys. Rev. D} \textbf{\bibinfo{volume}{97}},
  \bibinfo{pages}{023504} (\bibinfo{year}{2018}), \eprint{1709.02401}.

\bibitem[{\citenamefont{Fang et~al.}(2020)\citenamefont{Fang, Krause, Eifler,
  and MacCrann}}]{Fang++:2019}
\bibinfo{author}{\bibfnamefont{X.}~\bibnamefont{Fang}},
  \bibinfo{author}{\bibfnamefont{E.}~\bibnamefont{Krause}},
  \bibinfo{author}{\bibfnamefont{T.}~\bibnamefont{Eifler}}, \bibnamefont{and}
  \bibinfo{author}{\bibfnamefont{N.}~\bibnamefont{MacCrann}},
  \bibinfo{journal}{JCAP} \textbf{\bibinfo{volume}{05}}, \bibinfo{pages}{010}
  (\bibinfo{year}{2020}), \eprint{1911.11947}.

\bibitem[{\citenamefont{Chen et~al.}(2021)\citenamefont{Chen, Lee, and
  Dvorkin}}]{Chen++:2021}
\bibinfo{author}{\bibfnamefont{S.-F.} \bibnamefont{Chen}},
  \bibinfo{author}{\bibfnamefont{H.}~\bibnamefont{Lee}}, \bibnamefont{and}
  \bibinfo{author}{\bibfnamefont{C.}~\bibnamefont{Dvorkin}},
  \bibinfo{journal}{JCAP} \textbf{\bibinfo{volume}{05}}, \bibinfo{pages}{030}
  (\bibinfo{year}{2021}), \eprint{2103.01229}.

\bibitem[{\citenamefont{Feldbrugge}(2023)}]{Feldbrugge:2023}
\bibinfo{author}{\bibfnamefont{J.}~\bibnamefont{Feldbrugge}}
  (\bibinfo{year}{2023}), \eprint{2304.13064}.

\bibitem[{\citenamefont{Kaiser}(1992)}]{Kaiser:1991}
\bibinfo{author}{\bibfnamefont{N.}~\bibnamefont{Kaiser}},
  \bibinfo{journal}{Astrophys. J.} \textbf{\bibinfo{volume}{388}},
  \bibinfo{pages}{272} (\bibinfo{year}{1992}).

\bibitem[{\citenamefont{Kaiser}(1998)}]{Kaiser:1996}
\bibinfo{author}{\bibfnamefont{N.}~\bibnamefont{Kaiser}},
  \bibinfo{journal}{Astrophys. J.} \textbf{\bibinfo{volume}{498}},
  \bibinfo{pages}{26} (\bibinfo{year}{1998}), \eprint{astro-ph/9610120}.

\bibitem[{\citenamefont{LoVerde and Afshordi}(2008)}]{LoVerde:08}
\bibinfo{author}{\bibfnamefont{M.}~\bibnamefont{LoVerde}} \bibnamefont{and}
  \bibinfo{author}{\bibfnamefont{N.}~\bibnamefont{Afshordi}},
  \bibinfo{journal}{Phys. Rev. D} \textbf{\bibinfo{volume}{78}},
  \bibinfo{pages}{123506} (\bibinfo{year}{2008}), \eprint{0809.5112}.

\bibitem[{\citenamefont{Hamilton}(2000)}]{Hamilton:1999}
\bibinfo{author}{\bibfnamefont{A.~J.~S.} \bibnamefont{Hamilton}},
  \bibinfo{journal}{Mon. Not. Roy. Astron. Soc.}
  \textbf{\bibinfo{volume}{312}}, \bibinfo{pages}{257} (\bibinfo{year}{2000}),
  \eprint{astro-ph/9905191}.

\bibitem[{\citenamefont{Schmittfull et~al.}(2016)\citenamefont{Schmittfull,
  Vlah, and McDonald}}]{Schmittfull+:2016_I}
\bibinfo{author}{\bibfnamefont{M.}~\bibnamefont{Schmittfull}},
  \bibinfo{author}{\bibfnamefont{Z.}~\bibnamefont{Vlah}}, \bibnamefont{and}
  \bibinfo{author}{\bibfnamefont{P.}~\bibnamefont{McDonald}},
  \bibinfo{journal}{Phys. Rev. D} \textbf{\bibinfo{volume}{93}},
  \bibinfo{pages}{103528} (\bibinfo{year}{2016}), \eprint{1603.04405}.

\bibitem[{\citenamefont{Schmittfull and Vlah}(2016)}]{Schmittfull+:2016_II}
\bibinfo{author}{\bibfnamefont{M.}~\bibnamefont{Schmittfull}} \bibnamefont{and}
  \bibinfo{author}{\bibfnamefont{Z.}~\bibnamefont{Vlah}},
  \bibinfo{journal}{Phys. Rev. D} \textbf{\bibinfo{volume}{94}},
  \bibinfo{pages}{103530} (\bibinfo{year}{2016}), \eprint{1609.00349}.

\bibitem[{\citenamefont{McEwen et~al.}(2016)\citenamefont{McEwen, Fang, Hirata,
  and Blazek}}]{McEwen++:2016}
\bibinfo{author}{\bibfnamefont{J.~E.} \bibnamefont{McEwen}},
  \bibinfo{author}{\bibfnamefont{X.}~\bibnamefont{Fang}},
  \bibinfo{author}{\bibfnamefont{C.~M.} \bibnamefont{Hirata}},
  \bibnamefont{and} \bibinfo{author}{\bibfnamefont{J.~A.}
  \bibnamefont{Blazek}}, \bibinfo{journal}{JCAP} \textbf{\bibinfo{volume}{09}},
  \bibinfo{pages}{015} (\bibinfo{year}{2016}), \eprint{1603.04826}.

\bibitem[{\citenamefont{Simonovi\'c et~al.}(2018)\citenamefont{Simonovi\'c,
  Baldauf, Zaldarriaga, Carrasco, and Kollmeier}}]{Simonovic++:2017}
\bibinfo{author}{\bibfnamefont{M.}~\bibnamefont{Simonovi\'c}},
  \bibinfo{author}{\bibfnamefont{T.}~\bibnamefont{Baldauf}},
  \bibinfo{author}{\bibfnamefont{M.}~\bibnamefont{Zaldarriaga}},
  \bibinfo{author}{\bibfnamefont{J.~J.} \bibnamefont{Carrasco}},
  \bibnamefont{and} \bibinfo{author}{\bibfnamefont{J.~A.}
  \bibnamefont{Kollmeier}}, \bibinfo{journal}{JCAP}
  \textbf{\bibinfo{volume}{04}}, \bibinfo{pages}{030} (\bibinfo{year}{2018}),
  \eprint{1708.08130}.

\bibitem[{\citenamefont{Gao et~al.}(2023)\citenamefont{Gao, Vlah, and
  Challinor}}]{Gao++:2023}
\bibinfo{author}{\bibfnamefont{Z.}~\bibnamefont{Gao}},
  \bibinfo{author}{\bibfnamefont{Z.}~\bibnamefont{Vlah}}, \bibnamefont{and}
  \bibinfo{author}{\bibfnamefont{A.}~\bibnamefont{Challinor}}
  (\bibinfo{year}{2023}), \eprint{2305.xxxxx}.

\bibitem[{{\relax DLMF}()}]{NIST:DLMF}
{\relax DLMF}, \emph{\bibinfo{title}{{\it NIST Digital Library of Mathematical
  Functions}}}, \bibinfo{howpublished}{\url{https://dlmf.nist.gov/}, Release
  1.1.9 of 2023-03-15}, \bibinfo{note}{f.~W.~J. Olver, A.~B. {Olde Daalhuis},
  D.~W. Lozier, B.~I. Schneider, R.~F. Boisvert, C.~W. Clark, B.~R. Miller,
  B.~V. Saunders, H.~S. Cohl, and M.~A. McClain, eds.},
  \urlprefix\url{https://dlmf.nist.gov/}.

\end{thebibliography}

\end{document}